\documentclass[usenatbib]{mn2e}
\usepackage{float}
\usepackage{etex}
\usepackage{graphicx}
\usepackage{amsmath}
\usepackage{placeins}
\usepackage{threeparttable}
\usepackage{mathptmx}
\usepackage{txfonts}
\usepackage[T1]{fontenc}
\usepackage{ae,aecompl}
\usepackage{pdfpages}
\usepackage{etex}
\usepackage{tablefootnote}
\usepackage{scalerel}
\usepackage{longtable}
\usepackage{hyperref}
\usepackage{amsmath,amssymb}
\hypersetup{colorlinks, citecolor=blue}
\begin{document}

\title[How does dark matter stabilize disc galaxies? ]
{How does dark matter stabilize disc galaxies? }
\author[ K. Aditya]
        {K. Aditya \thanks{E-mail : aditya.k@iiap.res.in} \\
        Indian Institute of Astrophysics, Koramangala, Bengaluru 560 034, INDIA \\ 
} 
\maketitle

\begin{abstract}
The study presents a theoretical framework for understanding the role of dark matter on the stability of the galactic disc. We model the galaxy as a two-component system consisting of stars and gas in equilibrium with an external dark matter halo. We derive the equations governing the growth of perturbations and obtain a stability criterion that connects the potential of the dark matter halo and the gas fraction with the stability levels of the galaxy. We find that a two-component disc is more susceptible to the growth of gravitational instabilities than individual components, particularly as gas fractions increase. However, the external field, due to the dark matter halo, acts as a stabilizing agent and increases the net stability levels even in the presence of a cold gas component. We apply the stability criterion to models of the Milky Way, low surface brightness galaxies, and baryon-dominated cold rotating disc galaxies observed in the early universe. Our results show that the potential due to the dark matter halo plays a significant role in stabilizing nearby galaxies, such as the Milky Way, and low surface brightness galaxies, which would otherwise be prone to local gravitational instabilities. However, we find that the baryon-dominated cold disc galaxies observed in the early universe remain susceptible to the growth of local gravitational instabilities despite the stabilizing effect of the dark matter halo.
\end{abstract}
\section{Introduction}
Gravitational instabilities are fundamental processes that drive the evolution of the galaxy. It provides important clues for understanding how gas in the galaxies is converted into stars \citep{wang1994gravitational, dopita1994law, pandey1999gravitational,krumholz2010dynamics,forbes2014balance}, and 
how non-axisymmetric structures like bars and spiral arms form in galaxies \citep{goldreich1965ii,toomre1977theories, iye1978global, 
kalnajs1983theory,lin1987spiral, sellwood2000spiral,sellwood2012spiral}. One of the simplest diagnostics for accessing the stability of
the galactic disc against the growth of axisymmetric gravitational instabilities was proposed by \cite{toomre1964gravitational}. It measures the competing effect of self-gravity, which tries to destabilize the disc, and the stabilizing effect of the differential rotation and the random velocity dispersion. The balance between the 
stabilizing agents, i.e., differential rotation and random velocity dispersion, and the destabilizing agent, i.e., the self-gravity, is classically quantified by the stability criterion proposed by \cite{toomre1964gravitational}:
\begin{equation}
q=\frac{\kappa \sigma}{\pi G \Sigma}.
\end{equation}
In the above equation, $\kappa$ is the epicyclic frequency, $\Sigma$ is the mass surface density, and $\sigma$ is the radial velocity dispersion, where $q>1$ is the condition for stability of the disc against axisymmetric perturbations. The stability criterion proposed by \cite{toomre1964gravitational}  has been modified to include the self-gravity of both stars and gas by \cite{jog1984galactic, wang1994gravitational, elmegreen1995effective, jog1996local,rafikov2001local, romeo2011effective} and finally  \cite{romeo2013simple} derive a N-component stability parameter to quantify the stability of gravitationally coupled multiple stellar and gaseous discs. The stability parameter has also been modified further to include the physical processes like the effects of the turbulence \citep{2012MNRAS.425.1511H, 2015MNRAS.449.2156A} and the three-dimensional structure of ISM \citep{2022ApJ...937...88M}.\\
The two-component model for studying the stability of disc galaxy against the growth of local axisymmetrical gravitational instabilities was envisaged by \cite{jog1984galactic}. In the two-component model, stars and gas in the galactic disk are modeled as two isothermal fluids that interact gravitationally with each other. One of the components in the two-component model resembles the interstellar medium (ISM) with smaller values of the velocity dispersion, and the other resembles the stellar component with higher velocity dispersion. The approach has been used extensively to study the role of the cold ISM in driving the instabilities in galactic disc \citep{1984ApJ...276..127J} and for studying the stability of the Galactic disc by \citep{jog1996local}. Further, \cite{rafikov2001local} presents the stability criterion for a disc consisting of multiple isothermal components. Each component is categorized as either collisional, such as the ISM, or collisionless, such as stars. The results obtained by \cite{rafikov2001local} for the stability using a collisionless treatment for stars and collisional approach for the ISM are comparable to the results obtained \cite{jog1984galactic} and \cite{elmegreen1995effective}. The two-component stability parameter is a valuable diagnostic for understanding if the stability levels are driven by stars or by gas; for example, see \cite{romeo2017drives} and \cite{romeo2016powers}. \\
The stability criterion in the literature considers the self-gravity of the gas and stars. However, it does not consider the role of dark matter in driving the gravitational instabilities. The initial effort to incorporate the influence of a dark matter halo on the stability of a single-component disc was undertaken by \cite{jog2014effective}. In this work, we present the conditions for appraising the stability of the gravitationally coupled two-component disc consisting of stars and gas in equilibrium with an external dark matter halo. The differential equations governing the growth rate of perturbations are derived by considering the two-component disc in equilibrium with the external force field of the dark matter halo. Each component is specified by its velocity dispersion, surface density, and angular frequency, but the system is under the influence of the force field of the dark matter halo. We show that the governing equations for the growth of instabilities resemble a wave equation with extra terms. We use plane wave ansatz to derive the dispersion relation for the gravitationally coupled two-component system in equilibrium with the dark matter halo and obtain a simple stability criterion. The stability criterion presented in this work explicitly quantifies the contribution of dark matter to the overall stability levels. It can be used to explore the role of dark matter in regulating various physical processes within the galactic disc where gravitational instabilities are important.\\
The paper is organized as follows: in \S 2, we will formulate the basic equations and derive the governing differential equations. 
We will derive the dispersion relation and stability criterion in \S 3 and \S 4. We finally present the results in \S 5 and discuss the applications of the stability criterion in \S 6 $\&$ \S 7, and conclude in \S 8.

\section{Formulation and derivation of basic equations}
We consider a coaxial and coplanar thin disc comprising stars and gas, which interact with each other gravitationally. 
The two-component disc is supported by random pressure and rotation, and the system is in equilibrium with a constant 
external force field of the dark matter halo. The problem is described in the galactic cylindrical coordinate system 
$(R,\theta,z)$. We start with the basic hydrodynamic equations in which the external force field of 
the dark matter halo is in equilibrium with the two-component disc. We then introduce small perturbations in the basic equations and derive the dynamic equations governing the evolution of the perturbed quantities. The Force equation, continuity equation, and the Poissons equation for a thin disc in equilibrium with an external potential
$ \Phi_{ext} $ are:
\begin{equation}
\Sigma_{i} \frac{\partial \boldsymbol{V_{i}}}{\partial t} + \Sigma_{i}( \boldsymbol{V_{i}}.\boldsymbol{\nabla})\boldsymbol{V_{i}} = -\boldsymbol{\nabla} P_{i} - \Sigma_{i} \boldsymbol{\nabla} (\Phi_{s} +\Phi_{g} ) - \Sigma_{i} \boldsymbol{\nabla} \Phi_{ext},
\end{equation}

\begin{equation}
\frac{\partial \Sigma_{i}}{\partial t} +\boldsymbol{\nabla} .(\Sigma_{i} \boldsymbol{V_{i}})=0,
\end{equation}

\begin{equation}
\nabla^{2} (\Phi_{s} + \Phi_{g})=4\pi G(\Sigma_{s}+\Sigma_{g}) \delta(z).
\end{equation}
The above equations, when expressed in cylindrical coordinates, supplemented with an isothermal equation of state $ P_{i}= \Sigma_{i} c^{2}_{i}$ read:

\begin{equation}
\Sigma_{i}\frac{\partial u_{i}}{\partial t} + \Sigma_{i} u_{i} \frac{\partial u_{i}}{\partial R} + \frac{v_{i} \Sigma_{i}}{R} \frac{\partial u_{i}}{\partial \theta}
-\frac{\Sigma_{i} v^{2}_{i}}{R}=-c_{i}^{2} \frac{\partial \Sigma_{i}}{\partial R} -\Sigma_{i} \frac{\partial(\Phi_{s} + \Phi_{g} )}{\partial R} 
- \Sigma_{i} \frac{\partial \Phi_{ext}}{\partial R},
\end{equation}

\begin{equation}
\Sigma_{i} \frac{\partial v_{i}}{\partial t} + \Sigma_{i}u_{i} \frac{\partial v_{i}}{\partial R} + 
\frac{\Sigma_{i}v_{i}}{R} \frac{\partial v_{i}}{\partial \theta}  +\frac{\Sigma_{i} v_{i} u_{i}}{R} =-\frac{c_{i}^{2}}{R} \frac{\partial \Sigma_{i} }{\partial \theta} - \frac{\Sigma_{i}}{R} \frac{\partial (\Phi_{s}+\Phi_{g} )}{\partial \theta},
\end{equation}

\begin{equation}
\frac{\partial \Sigma_{i}}{\partial t} +\frac{1}{R} \frac{\partial(R u_{i} \Sigma_{i}) }{\partial R} + 
\frac{v_{i}}{R} \frac{\partial \Sigma_{i}}{\partial \theta} + \frac{\Sigma_{i}}{R}\frac{\partial v_{i}}{\partial \theta}=0,
\end{equation}

\begin{equation}
\frac{1}{R}\frac{\partial}{\partial R}\left( R \frac{\partial(\Phi_{s}+\Phi_{g} )}{\partial R}\right) +\frac{\partial^{2}(\Phi_{s} + \Phi_{g} ) }{\partial z^{2}}
 + \frac{1}{R^{2}} \frac{\partial^{2} (\Phi_{s} +\Phi_{g})}{\partial \theta ^{2}}=4 \pi G (\Sigma_{s} + \Sigma_{g}) \delta(z).
\end{equation}
In the above equations, $'i'$ is used to index $'stars'$ and $'gas'$, $ u_{i} $ and $ v_{i}$ are the velocity components in the 
radial and the tangential directions respectively, $ \Sigma_{i}$ and $ \Phi_{i}$ are the surface density and the 
gravitational potential associated with the stellar and the gas disc, respectively, and $c_{i}$ is the velocity dispersion of each 
component. Assuming the disc is axisymmetric, the above equations can be written as
\begin{equation}
\Sigma_{i}\frac{\partial u_{i}}{\partial t} + \Sigma_{i} u_{i} \frac{\partial u_{i}}{\partial R} 
-\frac{\Sigma_{i} v^{2}_{i}}{R}=-c_{i}^{2} \frac{\partial \Sigma_{i}}{\partial R} -\Sigma_{i} \frac{\partial(\Phi_{s} + \Phi_{g} )}{\partial R} 
- \Sigma_{i} \frac{\partial \Phi_{ext}}{\partial R},
\end{equation}

\begin{equation}
\Sigma_{i} \frac{\partial v_{i}}{\partial t} + \Sigma_{i}u_{i} \frac{\partial v_{i}}{\partial R} + 
\frac{\Sigma_{i} v_{i} u_{i}}{R} =0,
\end{equation}

\begin{equation}
\frac{\partial \Sigma_{i}}{\partial t} +\frac{1}{R} \frac{\partial(R u_{i} \Sigma_{i}) }{\partial R} =0,
\end{equation}

\begin{equation}
\frac{1}{R}\frac{\partial}{\partial R}\left( R \frac{\partial(\Phi_{s}+\Phi_{g} )}{\partial R}\right) +\frac{\partial^{2}(\Phi_{s} + \Phi_{g} ) }{\partial z^{2}}
 =4 \pi G (\Sigma_{s} + \Sigma_{g}) \delta(z).
\end{equation}
We now introduce small perturbations in the above basic equations:
\begin{equation}
\Sigma_{i} = \Sigma_{0,i} + \epsilon \Sigma_{1,i},\\
\Phi_{i} =\Phi_{0,i} + \epsilon \Phi_{1,i},\\
v_{i}=  v_{0,i} + \epsilon v_{1,i},\\
u_{i}= \epsilon u_{1,i}.
\end{equation}
The quantities $\Sigma_{0,i}$ and $\Phi_{0,i}, \,  v_{0,i}, \, u_{0,i}$  are the locally unperturbed states, and the  perturbed 
quantities are denoted by  $\Phi_{1,i}, \, \Sigma_{1,i}, \, v_{1,i}, \, u_{1,i}$, where the value of $\epsilon << 1$. 
Substituting equation (13) in equations  [(9), (10), (11), (12)] and keeping only the first order terms $[\epsilon^{1}]$, 
we obtain the governing equations for the perturbed quantities. But, before that, in 
order to better understand how the external potential interacts with the two-component $' star + gas'$ disc, we write down 
the zeroth order terms $[\epsilon^{(0)}]$ corresponding to equation (9):
 
\begin{equation}
   \frac{v^{2}_{0,i}}{R} = c^{2}_{i}\frac{\partial \ln \Sigma_{0,i}}{\partial R} +\frac{\partial \Phi_{0,s}}{\partial R} +\frac{\partial \Phi_{0,g}}{\partial R} 
     + \frac{\partial \Phi_{ext}}{\partial R}.
\end{equation}
In the above equation, the contribution of the term $c^{2}_{i}\partial \ln \Sigma_{0,i}/\partial R$ is negligible since the velocity dispersion is very small compared to 
the rotation velocity \citep{binney2011galactic}. Further, we write, $v^{2}_{0,s}=R\partial \Phi_{0,s}/\partial R$,  $v^{2}_{0,g}=R\partial \Phi_{0,g}/\partial R$ and $v^{2}_{ext}=R\partial \Phi_{ext}/\partial R$, we obtain: 
\begin{equation}
v^{2}_{net} = v^{2}_{0,s} + v^{2}_{0,g} + v^{2}_{ext}.
\end{equation}
 In the above equation, we have labeled $v_{0,i}$ as $v_{net}$ since it contains the effective contribution from the stars, gas, and the external potential. The value of $v_{net}$ is typically determined through observations of neutral hydrogen in galaxies \citep{de2001high,oh2011dark,lelli2016sparc}. We express the circular velocity of stars, gas, and the external potential in terms of 
the circular frequency as $v_{0,s}=R\Omega_{0,s}$, $v_{0,g}=R\Omega_{0,g}$ and $v_{ext}=R\Omega_{ext}$. 
This leads to $v^{2}_{net}= R^{2}\Omega^{2}_{net} =R^{2}(\Omega^{2}_{disc} + \Omega_{ext})$, where 
$\Omega^{2}_{disc}= \Omega^{2}_{0,s} + \Omega^{2}_{0,g}$.\\
The stability criterion derived by 
\cite{toomre1964gravitational}, \cite{goldreich1965ii}, \cite{jog1984galactic}, \cite{jog1996local} applies 
exclusively to star and/or gas disc. The net rotation contains the effect of only star and/or gas but does not
contain the contribution of the external potential to the net rotation, as shown in equation (15). 
In their treatment, the centrifugal force balances the unperturbed potential of either star or/and gas, i.e., $v^{2}_{net}/R =  \partial( \Phi_{0,s} + \Phi_{0,g} )/\partial R$ or $v^{2}_{net}/R =  \partial \Phi_{0} /\partial R$ for a single component,
but does not consider the contribution of the dark matter to the net rotation which enters our equations as an external potential 
$(\partial \Phi_{ext} /\partial R)$. However, when reconstructing stability using observed properties, the observed rotation curve is used, 
which includes contributions from stars, gas, and the dark matter halo. In contrast, the analytic treatment considers contributions only 
from the stars and/or gas disk. \cite{jog2014effective} identifies this difference between the analytical treatment and the observational
reconstruction of the stability criterion in the literature and derives a modified stability criterion for a one-component disc that includes 
the contribution of the dark matter halo to the net rotation.\\
Following the short detour aimed at understanding how the external potential interacts with the $'star+ gas'$ disc, we now write down the 
linearized equations governing the growth of perturbed quantities. The first order terms in $\epsilon^{(1)}$ are given by:

\begin{equation}
\frac{\partial u_{1,i} }{\partial t} -2 \Omega_{net} v_{1,i} + \frac{c_{i}^{2}}{\Sigma_{0,i}} \frac{\partial \Sigma_{1,i}}{\partial R} + 
\frac{\partial ( \Phi_{1,s} + \Phi_{1,g} )}{\partial R}  = 0,
\end{equation}

\begin{equation}
\frac{\partial v_{1,i} }{\partial t} -2B_{net} u_{1,i}=0,
\end{equation}

\begin{equation}
\frac{\partial \Sigma_{1,i} }{\partial t} + \Sigma_{0,i} \frac{\partial u_{1,i} }{\partial R}=0,
\end{equation}
and the Poisson equation for the thin disc assumes the form \citep{toomre1964gravitational};

\begin{equation}
\frac{\partial( \Phi_{1,s}+ \Phi_{1,g} )}{\partial R} = - 2 \pi i G( \Sigma_{1,s} + \Sigma_{1,g} ).
\end{equation}
In equation (16),  $ v_{net}$ is expressed as $v_{net}=\Omega_{net} R$, and the 
term $v^{2}_{net}/R -  \partial( \Phi_{0,s} + \Phi_{0,g} + \Phi_{ext} )/\partial R$ cancels, 
as it is just the centrifugal term balancing the total unperturbed potential of the two-component disc and the external potential.  
In equation (17), we have expressed $[\Omega_{net} +  \partial (\Omega_{net} R)/\partial R]  = -2B_{net}$, 
where $B_{net}$ is the Oort constant. Further, substituting $\Omega_{net}=\sqrt{\Omega^{2}_{disc} + \Omega^{2}_{ext}}$ 
in $[\Omega_{net} +  \partial (\Omega_{net} R)/\partial R]  = -2B_{net}$, it is straightforward\footnote{$\kappa^{2}_{net} =-4B_{net}\Omega_{net}=\left(R\frac{d\Omega^{2}_{disc}}{dR} + 4\Omega^{2}_{disc}\right) + \left(R\frac{d\Omega^{2}_{ext}}{dR} + 4\Omega^{2}_{ext}\right),\\ \kappa^{2}_{disc}=\left(R\frac{d\Omega^{2}_{disc}}{dR} + 4\Omega^{2}_{disc}\right), \,
\kappa^{2}_{ext}=\left(R\frac{d\Omega^{2}_{ext}}{dR} + 4\Omega^{2}_{ext}\right)$} to show that $\kappa^{2}_{net}=\kappa^{2}_{disc} + \kappa^{2}_{ext}$, where $\kappa_{net}$ is the net epicyclic frequency defined as $\kappa^{2}_{net}=-4B_{net}\Omega_{net}$. In equation (18), the term $ 1/R[ \partial(R u_{1,i} \Sigma_{0,i})/\partial R]$ is approximated as $\Sigma_{0,i}\partial u_{1,i}/ \partial R$, as $R\Sigma_{0,i}$ will vary gradually with $ R$ when compared with the rapid oscillatory behaviour of $u_{1,i}$.

\section{Dispersion relation in the presence of external field}
In this section, we will derive the dispersion relation for the two-component disc in the presence of an external field of the dark matter halo. We will show that the linearized equations [(16), (17), (18), (19)] governing the evolution of the perturbed quantities 
can be recast to resemble coupled wave equations with extra terms and thus admit solutions of the form $ e^{ik.r -\omega t}$.
Indexing equations [(16), (17), (18), (19)] for stars; \\
\begin{equation}
\frac{\partial u_{1,s}}{\partial t} -2 \Omega_{net} v_{1,s} + \frac{c_{s}^{2}}{\Sigma_{0,s}} \frac{\partial \Sigma_{1,s}}{\partial R} + 
\frac{\partial (\Phi_{1,s} +\Phi_{1,g} )}{\partial R}  = 0,
\end{equation}

\begin{equation}
\frac{\partial v_{1,s}}{\partial t} - 2B_{net}u_{1,s}=0,
\end{equation}

\begin{equation}
\frac{\partial \Sigma_{1,s}}{\partial t} + \Sigma_{0,s} \frac{\partial u_{1,s}}{\partial R}=0.
\end{equation}
Operating with $\partial/\partial R$ on equation (20), and eliminating the terms $\partial/\partial R \left(\partial u_{1,s}/\partial {t}\right)$ by taking the time derivative of equation (22), which will give $\partial/\partial R \left( \partial u_{1,s}/\partial {t} \right)= \left(-1/\Sigma_{0,s}\right)\partial^{2} \Sigma_{1,s}/\partial t^{2} $. 
Similarly, $\partial v_{1,s}/\partial R$ is eliminating by operating $\partial/\partial R$ on equation (21) and 
substituting for $\partial u_{1,s}\partial R$ from equation (22) to get $\partial v_{1,s}/\partial R =-2B_{net} \Sigma_{1,s}/\Sigma_{0,s}$. And finally substituting for $\partial^2 (\Phi_{1,s} +\Phi_{1,g} )/\partial R^2$ with equation (19), we obtain:

\begin{equation}
\frac{\partial^{2} \Sigma_{1,s}}{\partial t^{2}}  - c_{s}^{2} \frac{\partial^{2} \Sigma_{1,s}}{\partial R^{2}} -4\Omega_{net} B_{net} \Sigma_{1,s} + 
 2\pi i G \Sigma_{0,s}\frac{\partial}{\partial R}(\Sigma_{1,s} + \Sigma_{1,g}) =0.
\end{equation}
Similarly, the equation for gas reads ;
\begin{equation}
\frac{\partial^{2} \Sigma_{1,g}}{\partial t^{2}}  - c_{g}^{2} \frac{\partial^{2} \Sigma_{1,g}}{\partial R^{2}} -4\Omega_{net} B_{net} \Sigma_{1,g} + 
 2\pi i G \Sigma_{0,g}\frac{\partial}{\partial R}(\Sigma_{1,s} + \Sigma_{1,g}) =0.
\end{equation}
The above equations resemble wave equations and will indeed admit plane wave ansatz. Substituting $e^{i(k.r - \omega t)}$ for the perturbed quantities in equations (23) and (24), we obtain

\begin{equation}
\Sigma_{1,s} =\frac{-2 \pi G k \Sigma_{0,s} \Sigma_{1,g}}{(\omega^{2} -c_{s} ^{2} k^{2} -\kappa^{2}_{net} + 2 \pi G \Sigma_{0,s} k )},
\end{equation}
and similarly

\begin{equation}
\Sigma_{1,g} =\frac{-2 \pi G k \Sigma_{0,g} \Sigma_{1,s}}{(\omega^{2} -c_{g} ^{2} k^{2} -\kappa^{2}_{net} + 2 \pi G \Sigma_{0,g} k )}.
\end{equation}
 Combining equations (25) and (26), the final dispersion relation reads;

\begin{equation}
\begin{aligned}
(\omega^{2} -c_{s} ^{2} k^{2} -\kappa^{2}_{net} + 2 \pi G \Sigma_{0,s} k)
(\omega^{2} -c_{g} ^{2} k^{2} -\kappa^{2}_{net} + 2 \pi G \Sigma_{0,g} k)=\\
(2 \pi G \Sigma_{0,s} k)(2 \pi G \Sigma_{0,g} k).
\end{aligned}
\end{equation}
By setting the contribution of the external field to zero, i.e., $\kappa_{ext}=0$, equation (27) becomes equivalent to the dispersion relation for a two-component galactic disk, as shown in equation (17) of \cite{jog1984galactic}.
Further, if either of $\Sigma_{0,s},c_{s}=0$ or $\Sigma_{0,g},c_{g}=0$, equation (27) reduces to the case of a one-component disc under the influence of the external field \citep{jog2014effective}. In deriving the above dispersion relation, we have started with a two-component disc in equilibrium with an external force field of dark matter halo. We then introduced small perturbations and compose the linearized perturbation equations, which resemble plane wave equations and then 
use the plane wave ansatz to derive the dispersion relation. 

\section{Condition for stability}
In this section, we derive the stability criterion for assessing if 
the two-component disc in the force field of the dark matter halo is 
susceptible to the growth of axisymmetric instabilities or not.\\
Firstly, we define the following quantities:
\begin{equation}
\begin{aligned}
\alpha_{s}= \kappa^{2}_{net}  + c_{s} ^{2} k^{2} -  2 \pi G \Sigma_{0,s} k ,\\
\alpha_{g}= \kappa^{2}_{net}  + c_{g} ^{2} k^{2} -  2 \pi G \Sigma_{0,g} k ,\\
\beta_{s}=2 \pi G \Sigma_{0,s} k,\\
\beta_{g}=2 \pi G \Sigma_{0,g} k.
\end{aligned}
\end{equation}
Substituting equation (28) in (27), the dispersion relation and the respective roots are given by;
\begin{equation}
\begin{aligned}
\omega^{4} -\omega^{2}(\alpha_{s} + \alpha_{g})+(\alpha_{s} \alpha_{g} -\beta_{s} \beta_{g})=0\\
\omega^{2}_{\pm}=\frac{1}{2}(\alpha_{s} + \alpha_{g}) \pm \frac{1}{2}( (\alpha_{s}+\alpha_{g})^{2} -4(\alpha_{s} \alpha_{g} -\beta_{s} \beta_{g}))^\frac{1}{2}.
\end{aligned}
\end{equation}
For a one-component disc, $\alpha_{g} \geq 0$ or $\alpha_{s} \geq 0 $ is 
the sufficient condition for stability. For a marginally stable one-component disc a function $F$ 
can be defined as $ F= 2 \pi G \Sigma_{0} k/(\kappa^{2}_{net} + k^{2}c^{2} )$. A value of $F=1$ 
indicates marginal stability, $F>1$ represents an unstable disc and $F<1$ represents a stable disc.
The value of $k_{min}$ for the one-component disc is obtained by putting $d\omega^{2}/dk=0$, where 
$\omega^{2}= \kappa^{2}_{net} - 2 \pi G \Sigma_{0} k +c^{2}k^{2} $, which yields 
$k_{min} = \pi G \Sigma_{0}/c^{2}$. Evaluating $F$ at $k_{min}$ yields 
$F= 2/(1+Q^{2})$. For the one-component system in the force field of an external potential, Q is 
defined as $Q=\kappa_{net}c/\pi G\Sigma_{0}= q\sqrt{1+ (\kappa^{2}_{ext}/\kappa^{2}_{disc})}$, where $q=\kappa_{disc}c/\pi G \Sigma_{0}$.\\
The condition for marginal stability of two-component disc reads $\omega^{2}_{-}=0$ or $(\alpha_{s} \alpha_{g} -\beta_{s} \beta_{g})=0$,
and for the disc to be unstable the conditions is $\alpha_{s} \alpha_{g} -\beta_{s} \beta_{g} <0$.
With simple algebra the condition for neutral equilibrium, $(\alpha_{s} \alpha_{g} -\beta_{s} \beta_{g})=0$ can be written as:
\begin{equation}
F=\frac{2 \pi G \Sigma_{0,s} k}{\kappa^{2}_{disc} +\kappa^{2}_{ext} + k^{2}c_{s}^{2} } + \frac{2 \pi G \Sigma_{0,g} k}{\kappa^{2}_{disc} +\kappa^{2}_{ext}  + k^{2}c_{g}^{2} } 
\end{equation}
where F=1. In the above, we have expressed $\kappa^{2}_{net}= \kappa^{2}_{disc} +\kappa^{2}_{ext}$ to gauge the effect of the
external potential on the $'star + gas'$ disc. See, the discussion following equation (19) for deriving $\kappa_{net}$ in 
terms of $\kappa_{disc}$ and $\kappa_{ext}$. We define the gas fraction $f = \Sigma_{0,g}/(\Sigma_{0,s}+\Sigma_{0,g})$,  and
$X_{s-g}=\kappa^{2}_{disc}/[2 \pi G (\Sigma_{0,s}+\Sigma_{0,g})k_{min}]$. $X_{s-g}$ is the dimensionless wavelength at which it is 
hardest to stabilize the two-component system. The value of $k_{min}$ for the two-component system is given by conditions, $d\omega^{2}_{-}/dk=0$, or $d(\omega^{2}_{+}\omega^{2}_{-})/dk=0$, 
i.e. finding $d(\alpha_{s} \alpha_{g} -\beta_{s} \beta_{g})/dk$ which yields;
\begin{equation}
\begin{aligned}
k^{3}(4c_{s}^{2} c_{g}^{2}) - 3k^{2}(2\pi G\Sigma_{0,s} c_{g}^{2} +2\pi G\Sigma_{0,g} c_{s}^{2}) \\
+2k \kappa^{2}_{net}(c_{g}^{2}+c_{s}^{2})-(2\pi G\Sigma_{0,s}+2\pi G\Sigma_{0,g}) \kappa^{2}_{net}=0
\end{aligned}
\end{equation}
The function $F$ for the two-component model is a superposition of the one-component cases \citep{jog1984galactic}. Thus, in analogy with the one-component case, the condition for stability of the two-component disc under the force field of external potential is defined as:
\begin{equation}
\frac{2}{1+Q_{T}^{2}}=\frac{(1-f)}{X_{s-g}(1+ \frac{(1-f)^{2} q_{s}^{2}}{4X^{2}_{s-g}} + R )} +
			\frac{f}{X_{s-g}(1+ \frac{f^{2} q_{g}^{2}}{4X^{2}_{s-g}} + R  )}.
\end{equation}
In the above equation, $R$ quantifies the contribution of the external potential on the two-component $'star+gas'$ and is defined as $R=\kappa^{2}_{ext}/\kappa^{2}_{disc}$. Also, $q_{s}$ and $q_{g}$ are the classical one-component stability criterion for stars and gas, defined as $q_{s}=\kappa_{disc}c_{s}/\pi G\Sigma_{0,s}$ and $q_{g}=\kappa_{disc}c_{g}/\pi G\Sigma_{0,g}$ respectively.
The above condition is equivalent to the stability condition $Q_{s-g}$ derived by \cite{jog1996local} in the absence of the external force field $(R=0)$. For the sake of continuity of notation, we denote the stability criterion for the two-component disc in the absence of an external field using $q_{T}$.
The disc is stable against the growth of axisymmetric instabilities when $Q_{T}>1$, and the disc is susceptible 
to the growth of axisymmetric perturbations when $Q_{T}<1$. 

\section{Results}
\subsection{Marginal stability of one-component disc under the influence of dark matter halo}
To gain better insight into the role of dark matter on a two-component disc, we first investigate the impact of the force field 
of dark matter halo in driving the stability levels in a one-component disc. The dispersion relation for a one-component disc is given by:
\begin{equation}
\omega^{2}=(\kappa^{2}_{disc} + \kappa^{2}_{ext} )k^{(0)} + c^{2}k^{(2)} -2 \pi G \Sigma_{0} k^{(1)}.
\end{equation}
In the above equation, at a large value of $ k$,  $k^{2}$ will dominate; thus, pressure stabilizes the disc at small scales. At small $ k$, i.e., $ k^{(0)}$ the differential rotation of the disc ($\kappa^{2}_{disc}$) and the dark matter halo ($\kappa^{2}_{ext}$) stabilize the disc at large scales. At intermediate $k$, the self-gravity of the galactic disc becomes important. The field due to the external potential $(\kappa^{2}_{ext})$ adds up with the differential rotation of the disc $(\kappa^{2}_{disc})$ and will stabilize the disc. 
Next, we inspect the marginal stability of the one-component galactic disc. 
Putting $\omega^{2} =0$, equation (33) can be recast to obtain a quadratic equation in $k$, 
\begin{equation}
1 + \frac{Q^{2}}{4} \frac{k^{2}}{k'^{2}_{T}} -\frac{k}{k'_{T}}=0,
\end{equation}
where, $Q=q (1+ \kappa^{2}_{ext}/\kappa^{2}_{disc})^{\frac{1}{2}}$, $k'_{T}=k_{T}(1+\kappa^{2}_{ext}/\kappa^{2}_{disc})$, $k_{T}=\kappa^{2}_{disc}/2  \pi G \Sigma_{0}$ and 
defining $\zeta'= k'_{T}/k = k_{T}/k(1+\kappa^{2}_{ext}/\kappa^{2}_{disc})$, i.e $\zeta'=\zeta(1+ \kappa^{2}_{ext}/\kappa^{2}_{disc})$.
With the above substitutions, equation (34) can be written as $Q=2\left[ \zeta (1+R)\left( 1- \zeta (1+ R)  \right)\right]^{\frac{1}{2}}$, where $R=\kappa^{2}_{ext}/\kappa^{2}_{disc}$.
\begin{figure}
\centering
\resizebox{85mm}{75mm}{\includegraphics{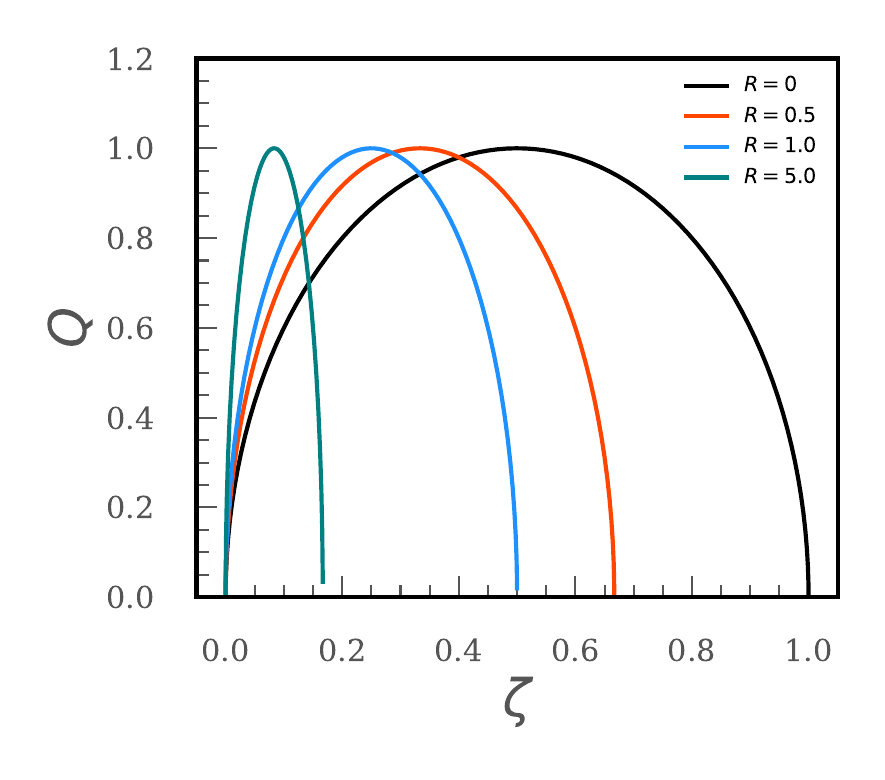}} 
\vspace{-11.0mm}
\caption{The marginal stability of a one-component disc under the influence of the external force field of dark matter halo.} 
\end{figure}
In Figure. 1, we show the effect of the external force field of the dark matter halo on the stability of the one-component disc. 
We find that upon increasing the contribution of the dark matter by increasing the value of $R$, the maximum value of $Q$ is shifted
towards a smaller value of $\zeta$, indicating that a larger contribution from dark matter to the total potential can effectively stabilize 
the galaxy over large scales. Further, from $Q=q \sqrt{1+ R}$, we can see that when $R=\kappa^{2}_{ext}/\kappa^{2}_{disc}=0$, 
the value of $Q$ corresponds to the classical stability criterion $(q)$ derived by \cite{toomre1964gravitational}. 
The stability criterion derived by \cite{toomre1964gravitational} considers the self-gravity of only one component and does not include the contribution of the external potential due to dark matter halo. The centrifugal force is balanced only by the corresponding force due to the unperturbed potential of stars/gas. The marginal stability in the absence of the external potential is given by $q=1$. 
The maximum value value of $Q$, when $R=0.5$, or $\kappa_{ext}=( \kappa_{disc}/\sqrt{2})$ is equal to 1.2 compared to 1 when R=0, indicating that $Q>q$. 
Thus, it is evident that the addition of dark matter to the total potential increases the marginal stability levels and makes it much harder to 
destabilize the one-component disc, making the disc more stable against the growth of instabilities. The stability criterion in the presence of the external dark matter halo 
$Q=q (1+ \kappa^{2}_{ext}/\kappa^{2}_{disc})^{\frac{1}{2}}$, can be written as $Q=\kappa_{net}c/\pi G\Sigma_{0}$. The one-component 
stability criterion derived by \cite{toomre1964gravitational} is applicable only for stars/gas. However, we note that the mathematical 
expression for the stability criterion in the presence of an external halo remains unchanged $(Q=\kappa_{net}c/\pi G\Sigma_{0})$. Thus, when 
reconstructing $q$ from observations following the classical treatment by \cite{toomre1964gravitational}, the contribution of the external force field of dark matter is implicitly accounted for. In other words, using net epicyclic frequency $(\kappa_{net})$ derived from the observed rotation curve in $q$ \citep{toomre1964gravitational} is equivalent to computing $Q$ derived in this work.

\subsection{Role of dark matter on the stability of two-component disc}
\begin{figure*}
\begin{center}
\resizebox{180mm}{115mm}{\includegraphics{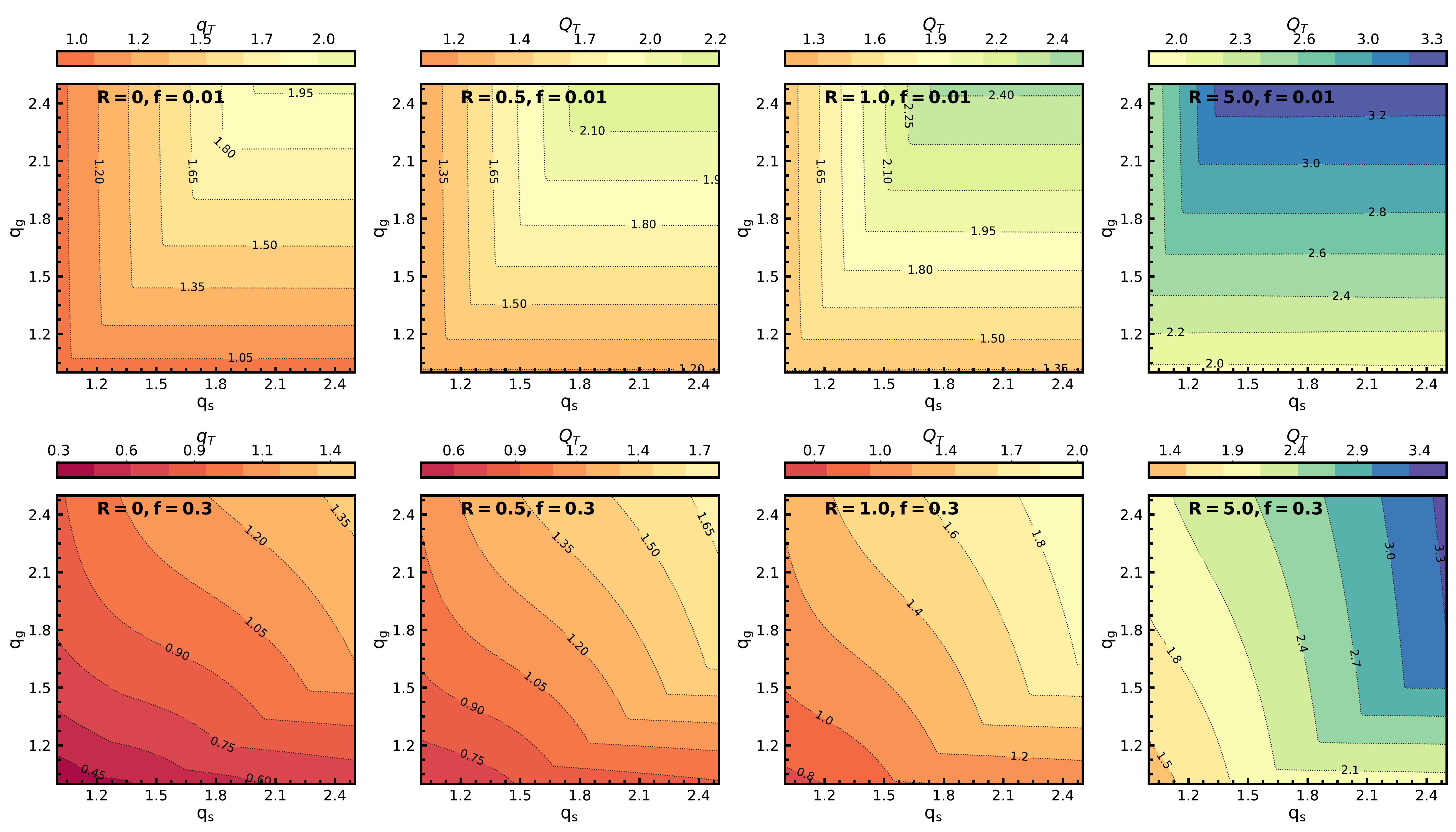}}
\end{center}
\caption{The effect of varying the external force of dark matter halo on the stability of the two-component disc. 
In the top panel, the gas fraction is fixed at f=0.01, and in the bottom panel, the gas fraction is fixed at f=0.3. The color bar indicates net stability levels given by $Q_{T}$.}
\end{figure*}
In \S 5.1, we found that the external potential due to the dark matter halo increases the marginal stability of the one-component disc, indicating that effectively $Q>q$, or that it is now harder to destabilize the disc due to the force field of the dark matter halo. \cite{jog1984galactic} and \cite{jog1996local}, show that the addition of gas disc makes the galaxy more prone to the growth of instabilities or in 
other words, the two-component disc is more unstable than either component by itself. The stability criterion presented in this 
work allows us to assess if gas is enough to lead to the growth of local instabilities, even in the presence of a stabilizing dark matter component. In Figure 2, we present the two-component stability criterion $Q_{T}$ as a function of $q_{s}$ and $q_{g}$, respectively. In the top panel, we have fixed the value of the gas fraction at $f=0.01$; in the bottom panel, we have fixed the gas fraction at 0.3. When the external force field due 
to the dark matter halo is zero $R=0$, we find that at a fixed value of the gas fraction, the value of the $q_{T}$ is lower than the values of
$q_{s}$ and $q_{g}$. This supports the earlier findings by \cite{jog1984galactic, jog1996local}, which show that a two-component disc is 
less stable than a disc composed only of stars or gas. The results indicate that the two-component disc is more prone to the development 
of gravitational instabilities than a single-component disc. For example, when $f=0.01, \, R=0$, value of $q_{T}=1.95$, when 
$q_{s}$ and $q_{g}$ are equal to 2.5. Similarly, when we increase the gas fraction to $f=0.3$, keeping $R=0$,
$q_{T}$ now becomes 1.35 when $q_{s},q_{g}=2.5$. Further, when $f = 0.3\, , R=0$, and $q_{s} \, ,q_{g}=1.5$, 
the value of $q_{T}$ drops to 0.75. This shows that adding a second disc in the absence of external potential due to dark matter 
effectively renders the two-component disc susceptible to the growth of gravitational instabilities, 
even though the stars and gas are stable by themselves.\\
We will now discuss the effect of dark matter on the stability of the two-component disc by varying the value of $R$. It is evident 
from Figure 2 that for a given gas fraction, when we move from left to right, the value of the marginal stability of the two-component
disc increases with increasing R. For example, at a gas fraction equal to 0.01, when both $q_{s}$ and $q_{g}$ are equal to 2.5, $q_{T}$ 
is equal to 1.95 for $R = 0$. However, upon increasing the contribution of dark matter to the total potential (i.e., $R = 1$), $Q_{T}$ 
becomes 2.4. A similar effect is observed at a higher gas fraction, when $f=0.3$ and both $q_{s}$ and $q_{g}$ are equal to 2.5, 
the value of $q_{T}$ is 1.35 in the absence of dark matter ($R=0$). However, when the contribution of dark matter is included ($R=1$), 
$Q_{T}=1.8$. The external potential of the dark matter halo stabilizes the two-component system, which would 
otherwise be prone to the growth of axisymmetric instabilities. At smaller values of $q_{s}$ 
and $q_{g}$ equal to 1.5 and a gas fraction equal to 0.3, the two-component disc becomes susceptible to the growth of gravitational instabilities 
($q_{T}=0.75$) when $R=0$. However, upon including the contribution of the dark matter halo $(R=1)$ at $f=0.3$, the two-component system stabilizes 
itself $(Q_{T}>1)$. Thus, we note that an external force due to the dark matter halo effectively suppresses the growth of local axisymmetric instabilities. However, the two-component system can be susceptible to axisymmetric instabilities 
on rarer occasions, even in the presence of stabilizing external potential due to the dark matter halo. An example is provided in Figure 2, when $R=0.5$, 
and the gas fraction is equal to $0.3$ and $q_{s},\,q_{g}\leq1.5$. The two-component system has $Q_{T}\leq1$, indicating that the system is prone to growth of local axisymmetric instabilities.

\section{Application}
\begin{figure*}
\begin{center}
\resizebox{180mm}{35mm}{\includegraphics{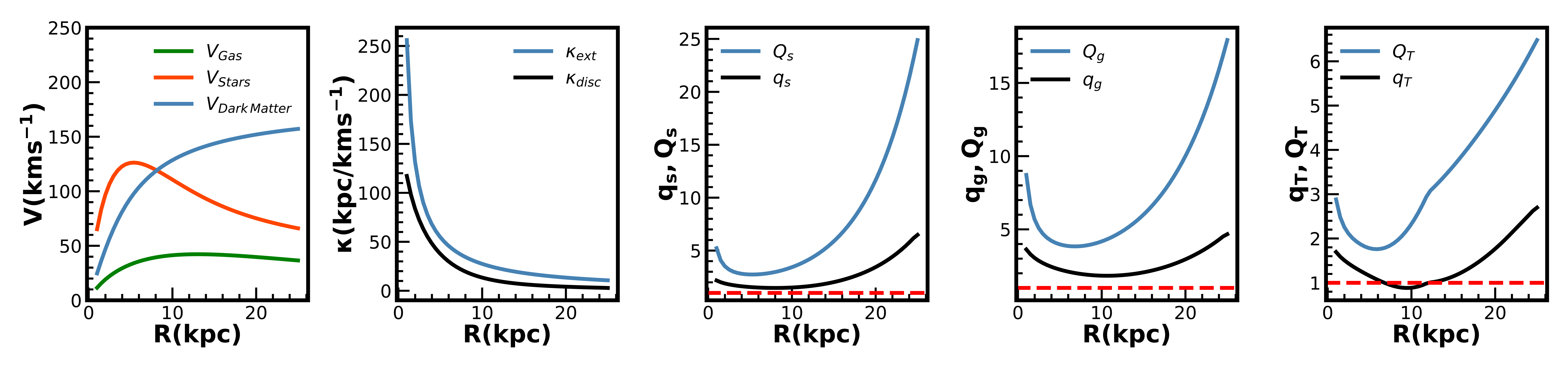}}\\
\resizebox{180mm}{35mm}{\includegraphics{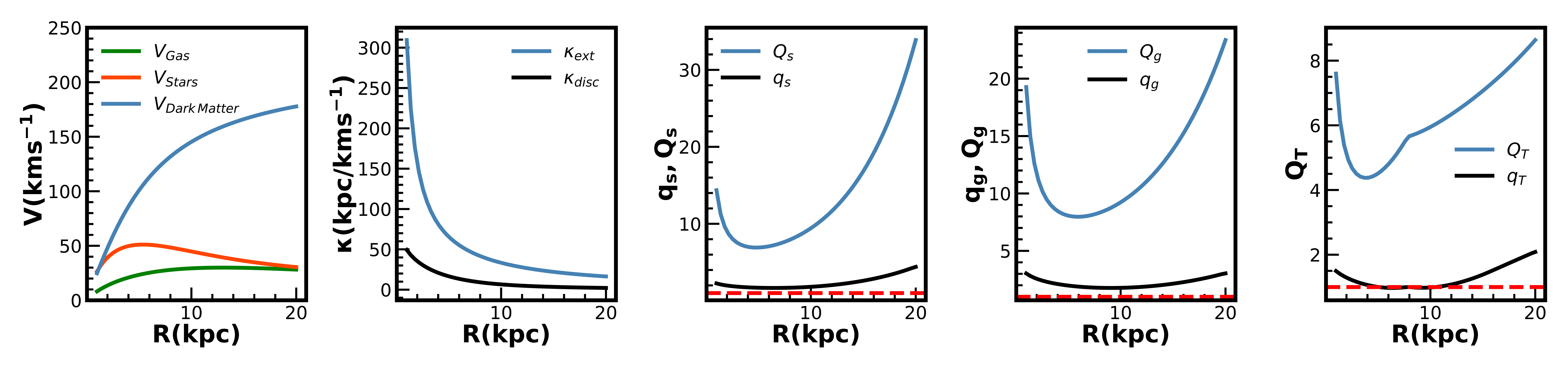}}\\
\resizebox{180mm}{35mm}{\includegraphics{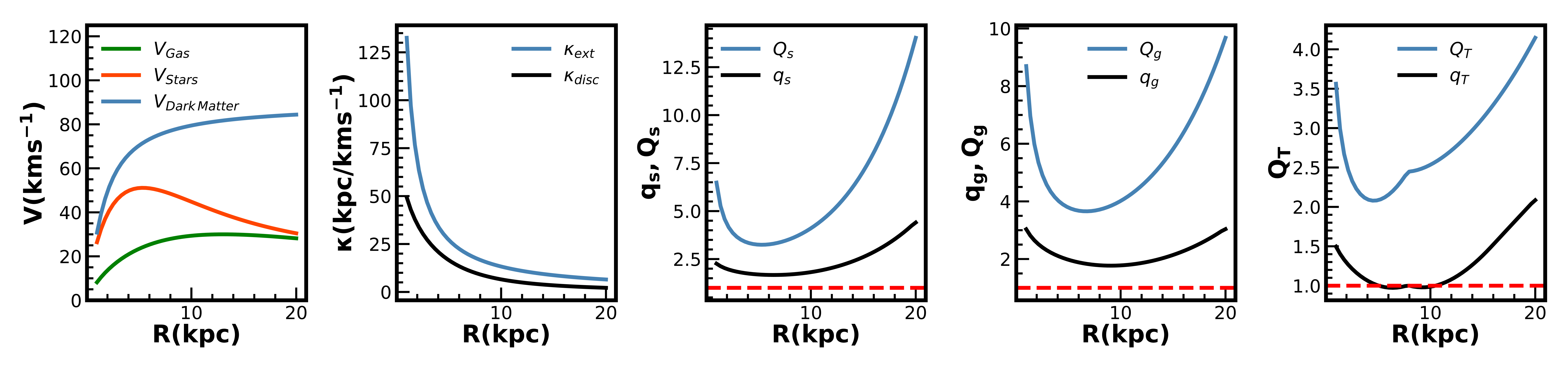}}\\
\resizebox{180mm}{35mm}{\includegraphics{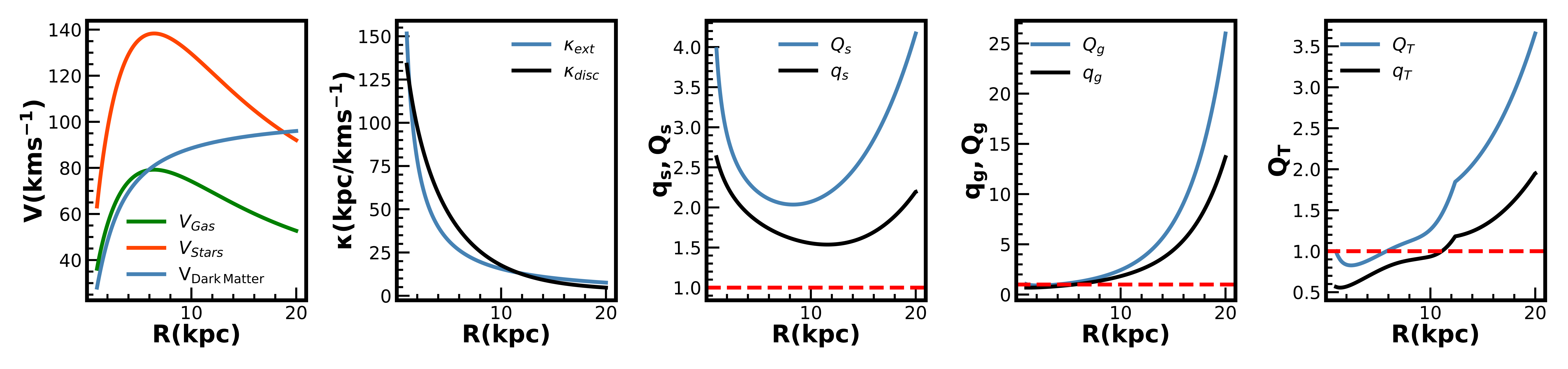}}\\
\end{center}
\caption{In the first row, we show stability analysis for model of Milky Way. In the second row, we show the stability analysis for a model of a low-mass stellar disc embedded in a massive dark matter halo. In the third row, we depict the analysis for a low-mass stellar disc embedded in a low-mass dark matter halo akin to a low surface brightness galaxy. The fourth panel shows the stability analysis for a baryon-dominated galaxy akin to cold rotating disc galaxies observed in the early universe. In realistic scenarios, $Q_{g}$, $Q_{s}$, and $Q_{T}$ correspond to the stability curves estimated from observations. The red dashed line indicates marginal stability levels.}
\end{figure*}

From the analysis presented in the previous section, we understand that the external force field of the dark matter halo stabilizes the 
two-component system of $stars+gas$. However, in rare instances where the force due to the dark matter halo is insufficient compared to 
the destabilizing effect of the gas disc, the two-component system may become prone to the growth of local axisymmetric instabilities. 
In this section, we will investigate the role of dark matter on the stability of two-component models of nearby galaxies 
like the Milky Way and low surface brightness galaxies, and models of galaxies observed in the early universe. 
The stellar distribution in our galaxy follows an exponential surface density given by
\begin{equation}
\Sigma_{s}(R)=\Sigma_{s0}e^{-R/R_{D}},    
\end{equation}
where $\Sigma_{s0}=640M_{\odot}pc^{-2}$ and $R_{D}=3.2kpc$ \citep{mera1998towards} are the central surface density and the disc scalelength.
The gas distribution in the Galaxy is given by 
\begin{equation}
    \Sigma_{g}(R)=\Sigma_{g0}e^{-1.65R/R_{25}},    
\end{equation}
in the above equation $\Sigma_{g0}=28.2M_{\odot}pc^{-2}$ is the central density of the gas disc and 
$R_{25}$ is the radius at which the B-band surface brightness drops to $25.5\, mag \, arsec^{-2}$, $R_{25}=4R_{D}$ \citep{bigiel2012universal}. 
The stellar velocity dispersion is given by \citep{leroy2008star,romeo2017drives} $\sigma_{s}(R)= (1/0.6) \sqrt{ (2 \pi G R_{D}\Sigma_{s}(R)/7.3}$, and we use a constant gas velocity dispersion equal to  $10kms^{-1}$ \citep{tamburro2009driving, mogotsi2016hi}. 
The circular velocity $(v_{c})$ corresponding to the exponential distribution is given as \citep{binney2011galactic} 
\begin{equation}
  v^{2}_{c}(R)=  4\pi G\Sigma_{0}R_{D}y^{2}[ I_{0}(y)K_{0}(y) -I_{1}(y)K_{1}(y)],
\end{equation}
where $y=R/2R_{D}$ and $I_{0},I_{1}$ and  $K_{0},K_{1}$ are the modified Bessel functions of the first and second kind. 
The epicyclic frequency $\kappa$ at a radius R is defined as 
\begin{equation}
\kappa^2(R)= \bigg( R\frac{d\Omega^{2} (R)}{dR} + 4\Omega^{2} (R)  \bigg),
\end{equation}
where $\Omega$ is the angular frequency defined as $\Omega^{2} (R)= \frac{v^{2}_{c}}{R^{2}}$. 
The dark matter density is given by a pseudo-isothermal halo, with a central density $\rho_{0}=0.035M_{\odot}pc^{-3}$ 
and a core radius $R_{c}=5kpc$ \citep{mera1998towards}. The epicyclic frequency due to the pseudo-isothermal dark matter halo is given by
\begin{equation}
\kappa^{2}_{PIS}(R)= 4 \pi G \rho_{0}\left[ \frac{2 R_c^2}{R^2 + R_c^2} + \frac{R_c^4}{R^2 (R^2 + R_c^2)} -   \frac{R_c^3}{R^3} \tan^{-1} \left( \frac{R}{R_c} \right)                \right].
\end{equation}
With all the building blocks needed to compute the stability in place, we will now discuss the role of dark matter halo on different galaxy models.\\
\textbf{Case 1: Stability of Milky Way}\\
We show the stability analysis of the Milky Way in the first row of Figure 3. We find the minimum value of the stability for stars and gas is $q^{min}_{s}=1.5$ and $q^{min}_{g}=1.8$, indicating that individually stars and gas are stable. However, the two-component formalism yields a $q^{min}_{T}=0.88$,  indicating that the two-component system is prone to the growth of local gravitational instabilities in the absence of the external potential due to the dark matter halo. Now, the addition of the dark matter to the total potential naturally increases the one-component stability from $q^{min}_{s}=1.5$ to $Q^{min}_{s}=2.7$ and $q^{min}_{g}=1.8$ to $Q^{min}_{g}=3.8$. Moreover, the two-component system, which was unstable with $q^{min}_{T}=0.88$, now has $Q^{min}_{T}=1.7$, highlighting the importance of the external potential of the dark matter halo in stabilizing massive disc galaxy like Milky Way.\\
\textbf{Case 2: Low mass disc in a high mass halo}\\
In order to better gauge the role of dark matter on the stability levels of the two-component disc, we minimize the contribution of the stars and gas disc to the total potential. We lower the stellar and the gas surface density to typical values observed in the low surface brightness galaxies; $\Sigma_{s0}=100M_{\odot}pc^{-2}$ and $R_{D}=2.5kpc$ and the gas surface density to $14.2M_{\odot}pc^{-2}$ \citep{de2001high,di2019universal}.
We keep the values of the dark matter halo to that of the Milky Way. We show the rotation velocity for 
this mass distribution in the second row of Figure 3. We can see that $\kappa_{ext}$ is significantly 
higher than $\kappa_{disc}$, highlighting that the dark matter is the dominant mass component. We find 
that $q^{min}_{s} = 1.7$ and $q^{min}_{g} = 1.8$, indicating that the stars and gas are stable on their own. 
However, similar to the Milky Way, the two-component star+gas system has $q^{min}_{T} = 0.9$, making the disc 
susceptible to the growth of local gravitational instabilities in the absence of the dark matter halo. 
However, upon including the contribution of the dark matter halo, we find $Q^{min}_{s} = 6.9$, $Q^{min}_{g} = 8$, 
and $Q^{min}_{T}=4.4$, indicating  that a higher contribution of the dark matter to the total potential is 
reflected in higher net stability levels of the two-component system.\\
\textbf{Case 3: Low mass disc in a low mass halo}\\
As a final example, we will inspect the effect of a low-mass stellar and gas disc embedded in a low-mass dark 
matter halo akin to the mass distribution of a low surface brightness galaxy. We keep the surface density of the 
stars, gas, and dark matter halo parameters to the typical values obtained from the mass models of low 
surface brightness galaxies: $\rho_{0}= 0.066M_{\odot}pc^{-3}$ and $R_{c}=1.5kpc$ \citep{de2008mass, di2019universal}. 
The parameters for the stars and gas are the same as in Case 2. We show the rotation velocity corresponding to this 
mass distribution in the third row of Figure 3. The minimum value of $q_{s}$, $q_{g}$ and $q_{T}$ are comparable to 
values obtained in case 2: $q^{min}_{s}=1.7$, $q_{g}=1.8$ and $q_{T}=0.9$. However, since the contribution of dark matter 
to the total potential is small compared to the massive dark matter halo of the Milky Way, the shift in the stability curves 
upon adding a dark matter halo is also small. We find that $Q^{min}_{s}=3.2$, $Q^{min}_{g}=3.6$ and $Q^{min}_{T}=2.1$. 
In both Case 2 and Case 3, we find that the force due to the dark matter potential stabilizes the two-component low surface brightness 
disc, which is otherwise unstable. The only difference is that a massive halo contributes significantly to the overall stability. 
This aligns with the previous finding in \cite{garg2017origin, aditya2023stability}, which shows that dark matter is important 
in regulating the stability of low surface brightness galaxies.

\section{Discussion}
A large number of recent studies show that the galaxies observed at high redshift are dominated by baryons 
\citep{rizzo2021dynamical, genzel2020rotation, genzel2017strongly, genzel2014sins}. We construct a galaxy 
model in which the contribution of stars and gas exceeds that of the dark matter halo in the total mass budget.
The rotation curve decomposition by \cite{rizzo2021dynamical} show that stellar disc makes maximum contribution 
to the total rotation curve. However, the contribution from the 
dark matter halo and the gas disc are typically comparable. In our model, the stellar 
disc has a surface density profile comparable to the Milky Way: $\Sigma_{0} =640M_{\odot}pc^{-2}$ and $R_{d}=3.2kpc$. 
We increase the gas surface density from $28.2M_{\odot}pc^{-2}$ for the Milky Way to  $200M_{\odot}pc^{-2}$ and 
keep the scalelength comparable to the stellar disc. However, the gas disc continues to be a cold component with a velocity
dispersion of $10 km/s$. We also lower the contribution of the dark matter by reducing the dark matter density and 
core radius to $0.05M_{\odot}pc^{-3}$ and $2kpc$, respectively. We aim to ascertain the contribution of the dark matter 
halo to the net stability levels in baryon-dominated systems akin to the cold rotating disc galaxies observed in the early universe. 
We show the results in the fourth panel of Figure 3. We find that in the absence of potential due to the dark matter halo, $q^{min}_{s}=1.5$, $q^{min}_{g}=0.8$ and $q^{min}_{T}=0.5$, indicating that a massive cold two-component disc is susceptible to the growth of local gravitational instabilities. 
Although the dark matter halo increases the net stability levels, the two-component system is still susceptible to local 
gravitational instabilities \citep{bacchini20243d, aditya2023stability}, $Q^{min}_{s}=2.8$, $Q^{min}_{g}=0.9$, and $Q^{min}_{T}=0.8$. 
This indicates that despite the stabilizing nature of the dark matter halo, the net contribution of the dark matter is 
insufficient to stabilize the baryon-dominated cold disc galaxies in the early universe.

\section{Conclusions}
In this study, we have derived detailed theoretical formalism to understand the role of dark matter 
and gas fraction on the stability of the two-component model of galactic disc. We model the galaxy as 
a coplanar and a coaxial system of stars and gas in equilibrium with an external dark matter halo. 
We derive the equations governing the growth rate of perturbation and, finally, present a simple 
stability criterion for appraising the stability of the two-component disc under the influence of 
dark matter halo. We find that:
\begin{enumerate}
\item The two-component disc is more susceptible to the growth of gravitational instabilities than the individual components. 
Increasing the gas fraction at a fixed value of external potential lowers the stability of the two-component disc, 
highlighting the role of cold gas in destabilizing the galaxy consistent with the earlier finding of \cite{jog1984galactic}.\\

\item The external field due to the dark matter halo acts as a stabilizing agent and increases the net stability levels of the 
two-component system. In dark matter-dominated systems, the gravitational force exerted by the dark matter halo stabilizes 
the two-component system, even when the system is locally unstable  \citep{jog2014effective}. This indicates that the cold gas 
component cannot destabilize the two-component disc when the dark matter halo dominates the mass budget of the galaxies.\\

\item We apply the stability criterion to the models of the Milky Way and low surface brightness galaxies 
and find that the Milky Way and the low surface brightness discs are locally unstable, when contribution of 
dark matter is not included in the total potential. However, the addition of the dark matter to the total potential 
significantly increases the net stability levels in these galaxies \citep{aditya2023stability}. We note that when the contribution 
of dark matter to the total mass budget is small, the corresponding effect on the net stability levels would also be diminished.\\

\item In rare cases, the two-component system can be susceptible to the growth of gravitational instabilities 
despite the presence of a stabilizing dark matter halo potential. One example is found in baryon-dominated  
cold rotating disc galaxies observed in the early universe. The influence of dark matter on the overall gravitational 
potential is insufficient to stabilize the galaxies observed in early universe.\\

\end{enumerate}

\section{acknowledgements}
Aditya would like to thank the referee for their insightful comments that improved the quality of this manuscript.

\section{Data Availability}
No new data was generated in this work.

\small{\bibliographystyle{mnras}}
\bibliography{example} 

\end{document}